\documentclass[12pt]{article}%
\usepackage{amsfonts}
\usepackage{amsmath}
\usepackage{geometry}
\usepackage{amssymb}
\usepackage{graphicx}
\usepackage[onehalfspacing]{setspace}
\usepackage{hyperref}
\usepackage[nosort]{cite}
\usepackage{appendix}%
\providecommand{\U}[1]{\protect\rule{.1in}{.1in}}
\begin{document}

\title{\textbf{Mental Age Compatibility: Quantification through the Convolution of
Probability Distributions}}
\author{\textbf{Patrick A. Haas\medskip}\\Department of Physics\\California State University, Dominguez Hills\\Carson, CA 90747, USA\\\medskip\\phaas@csudh.edu}
\date{}
\maketitle

\begin{abstract}
\noindent\thispagestyle{empty}We build on the empirical finding that a human
being's mental age is normally distributed around the chronological age. This
opposes the frequent societal assumption \textquotedblleft mental =
chronological\textquotedblright\ which is known to be false in general but
entertained for simplicity due to lack of methodology; hence disregarding
that, f.e., people of different chronological ages can be much closer in their
mental ages. As a quantitative approach on a scientific basis, we set up a
general formula for the probability that two individuals of given ages are
mentally within a certain range of years and investigate its implications i.a.
by critically analyzing popular assumptions on age and computing statistical
expectations within populations.\newpage\setcounter{page}{1}

\end{abstract}
\tableofcontents

\section{Introduction}

It is very well known that human beings do not behave, think, and feel across
the board according to the stereotypes assigned to their chronological ages.
Statements like \textquotedblleft he does not act like someone his
age\textquotedblright\ or \textquotedblleft she is far beyond her
age\textquotedblright\ are heard often throughout different societies. The
reality behind this and related phenomena is usually described within the
context of mental age.

The fact that people can be mentally older or younger than what is usually
referred to as their \textquotedblleft age\textquotedblright, raises the
question how far two individuals are actually apart in age on a mental level,
which necessarily leads to questioning a lot of profound (legal, moral,
ethical) views regarding \textquotedblleft age disparities\textquotedblright%
\ anchored in common thinking and legislation.

In many societal issues, it is assumed that mental age is equal to
chronological age. Of course, people are aware that in general these two ages
cannot be equal; however, in the absence of a reliable calibration, such an
assumption is nonetheless made for simplicity.

In this work, we propose a statistical method that promises to shed light on
the issue and might provide guidance for practical applications. Although a
statistical estimation cannot provide a conclusive answer to what is right or
wrong in a particular case, it can furnish an enlightening -- and quantitative
-- measure for good reasoning to reconsider accustomed ways of thinking and
judging and so raise awareness of reality being in many cases and regards
likely significantly different than commonly presupposed through socially
conditioned mindsets.

The general concept of mental age is a measure of (emotional) intelligence and
hence a rather multifaceted quantity across a spectrum of various cognitive
abilities; e.g. the five components of emotional intelligence by Goleman
\cite{Goleman}: self-awareness, self-regulation, social skills, empathy,
motivation; the four factors of the Emotional Intelligence Inventory (EII) by
Mayer and Salovey \cite{MayerSalovey1,Tapia} (analyzed for its correlation to
the Emotional Intelligence Scales (EIS) \cite{SchutteMalouffHall} by Tapia and
Marsh II \cite{TapiaMarshII}): empathy, utilization of feelings, handling
relationships, and self-control. That is why the mental age of an individual
is not unique but dependent on the considered modality of intelligence and its
components. It has also been subject to debate among groups of scholars, since
there has been dissent about whether intelligence is permanently ingrained
into someone's DNA or variable through external factors of influence.

A number of studies and tests have been conducted on the topic, which obtained
prolific results by focussing on specific or various aspects of interest. In
particular, special point schemes have been designed for that purpose, on the
basis of which subjects were tested. According to the Binet-Simon Intelligence
Scales (IQ test) \cite{Binet-Simon}, later revised by Terman et Al.
\cite{Terman1,Terman2}, scores were separately averaged by age groups and
those averages used as benchmarks for the corresponding mental age. Regardless
of the details (gender, culture, the modality and its components in question
etc.), a broad finding was that the scores from manifold tests aimed at
measuring intelligence are normally (Gaussian) distributed around the found
average values of various chronological age groups. This makes intuitive and
logical sense: people are more likely to be mentally closer to the average of
their chronological age cohort; yet, even if they experience similar social
structures and cultural habits and evolve through connatural forms of mental
development within the same amount of time, there is a multiplicity of factors
that give rise to individual divergences, that is, statistical deviations from
the average. Given a mean value and a standard deviation, the principle of
maximum entropy states a Gaussian form for the corresponding probability distribution.

It is important to note that the minutiae on the concept of intelligence and
its in-depth measurement process are not relevant for the purpose of the
current work. We centralize the empirical observation based on numerous
studies that (emotional) intelligence is quantifiable, particularly assignable
to age, and that the underlying mathematical structure for its distribution --
and hence the distribution of mental age around a given chronological age --
is a Gaussian probability density function (pdf).

Mental age is predominantly regarded to emotional intelligence here. From
studies elaborating on its validation and estimation (e.g.
\cite{SchutteMalouffHall,TapiaMarshII,Bar-On,EQYoungAdults,EQOlderChildren,Amantha}%
), with age groups ranging from\ older children (10-13 years) to young adults
(18-29 years), we can distill a quite compact scope of guide values for the
standard deviation.

In section 2, we set up the pdf for mental age and, in virtue of the
convolution formalism, derive a formula for the probability that two people of
given chronological ages are within a certain mental age range. We also set up
probability formulae for further specific situations.

In section 3, we derive formulae for statistical expectations based on the
aforementioned probability.

In section 4, we discuss ranges for the parameters, like the standard
deviation, by reasoning natural bounds, applying empirical data from studies,
and analyzing the mathematical structure.

In section 5, we elaborate in depth on the implications of the presented
formalism, i.a. by computing explicit numerical results for given groups of people.

In the appendix, we provide an error analysis, accounting for the negligible
impact of necessary approximations and the propagation of errors of parameters
for the probability value.

\section{Probability}

\subsection{General case}

For a randomly chosen individual's chronological age, $\mu$, we assume a
normal distribution for the spectrum of possible mental ages, $x$,%
\begin{equation}
g\left(  x;\mu,\sigma^{2}\right)  =\tfrac{1}{\sqrt{2\pi}\sigma}e^{-\frac{1}%
{2}\left(  \frac{x-\mu}{\sigma}\right)  ^{2}}, \label{Gauss}%
\end{equation}
where $\sigma$ is the standard deviation, and ages are given in years
througout this work.

Note: Very strictly spoken, we would need to consider a truncated Gaussian pdf
for the mental age distribution, with a lower bound within the period
$x_{0}\in\left[  -0.75,0\right]  $, since that is the point where a human
being starts to develop ($x=0$ set at birth), and an upper bound at some
higher age; however, the contributions of outer Gaussian sectors, especially
in case of the product of two Gaussians, are negligibly small, as we show in
the error discussion in the appendix.

If we randomly choose two people of ages $\mu_{1}$ and $\mu_{2}$,
respectively, what is the probability that their mental ages are lying at most
$d$ years apart? Going with $\left(  \ref{Gauss}\right)  $, the probability of
person 1 having their mental age within $\left[  x_{1},x_{1}+dx_{1}\right]  $
is $g_{1}\left(  x_{1};\mu_{1},\sigma_{1}\right)  dx_{1}$ and person 2 being
within the range $\left[  x_{1}-d,x_{1}+d\right]  $ is $\int_{x_{1}-d}%
^{x_{1}+d}g_{2}\left(  x_{2};\mu_{2},\sigma_{2}^{2}\right)  dx_{2}$, so that
the sought result can be very well approximated by taking the integral of the
product of these expressions:%
\begin{equation}
p\left(  d;\mu_{1},\sigma_{1},\mu_{2},\sigma_{2}\right)  =\int_{-\infty
}^{\infty}dx_{1}\int_{x_{1}-d}^{x_{1}+d}dx_{2}g_{1}\left(  x_{1};\mu
_{1},\sigma_{1}^{2}\right)  g_{2}\left(  x_{2};\mu_{2},\sigma_{2}^{2}\right)
. \label{P_Integral}%
\end{equation}

This can also be expressed through the convolution of $g_{1}$ and $g_{2}$,
where $\mu_{2}$ needs to be given the opposite sign\footnote{In fact, it does
not matter which $\mu$ gets the negative sign; the idea is to imply a pdf for
$\Delta\mu$ by the convolution.}, integrated over the mental age span:%
\begin{equation}
p\left(  d;\mu_{1},\sigma_{1},\mu_{2},\sigma_{2}\right)  =\int_{-d}%
^{d}dy\left(  g_{1}\ast g_{2}\right)  \left(  y;\mu_{1},\sigma_{1}^{2}%
,-\mu_{2},\sigma_{2}^{2}\right)  . \label{Convolution}%
\end{equation}

Applying%
\begin{equation}
\mathcal{N}\left(  \mu_{1},\sigma_{1}^{2}\right)  \ast\mathcal{N}\left(
-\mu_{2},\sigma_{2}^{2}\right)  =\mathcal{N}\left(  \mu_{1}-\mu_{2},\sigma
_{1}^{2}+\sigma_{2}^{2}\right)
\end{equation}
to $\left(  \ref{Gauss}\right)  $, and using the cumulative distribution
function (cdf),%
\begin{equation}
\Phi\left(  \tfrac{x-\mu}{\sigma}\right)  =\tfrac{1}{\sqrt{2\pi}\sigma}%
\int_{-\infty}^{x}dye^{-\frac{1}{2}\left(  \frac{y-\mu}{\sigma}\right)  ^{2}}%
\end{equation}
we can solve $\left(  \ref{Convolution}\right)  $ to%
\begin{equation}
p\left(  d;\mu_{1},\sigma_{1},\mu_{2},\sigma_{2}\right)  =\Phi\left(
\tfrac{\mu_{1}-\mu_{2}+d}{\sqrt{\sigma_{1}^{2}+\sigma_{2}^{2}}}\right)
-\Phi\left(  \tfrac{\mu_{1}-\mu_{2}-d}{\sqrt{\sigma_{1}^{2}+\sigma_{2}^{2}}%
}\right)  . \label{P_Result}%
\end{equation}

\subsection{Special case}

Assuming that the mental age of one person is known to be $x_{1}$, the
probability of them to be mentally $d$ or less years away from another
randomly chosen person with pdf $g\left(  x;\mu,\sigma^{2}\right)  =\tfrac
{1}{\sqrt{2\pi}\sigma}e^{-\frac{1}{2}\left(  \frac{x-\mu}{\sigma}\right)
^{2}}$ is computed by%
\begin{equation}
p\left(  d,x_{1};\mu,\sigma\right)  =\int_{x_{1}-d}^{x_{1}+d}dxg\left(
x;\mu,\sigma^{2}\right)  =\Phi\left(  \tfrac{x_{1}-\mu+d}{\sigma}\right)
-\Phi\left(  \tfrac{x_{1}-\mu-d}{\sigma}\right)  ,
\end{equation}
which can be easily obtained by using $\left(  \ref{P_Integral}\right)  $ and
setting the pdf of the person of known mental age equal to a Dirac delta
function with peak at $x=x_{1}$.

\subsection{Normalized case}

Given two randomly chosen individuals of ages $\mu_{2}\leq\mu_{1}$, one can
normalize the probability $\left(  \ref{P_Result}\right)  $ as per%
\begin{equation}
p_{0}\left(  d;\mu_{1},\sigma_{1},\mu_{2},\sigma_{2}\right)  =\tfrac{p\left(
d;\mu_{1},\sigma_{1},\mu_{2},\sigma_{2}\right)  }{p\left(  d;\mu_{2}%
,\sigma_{2},\mu_{2},\sigma_{2}\right)  }, \label{P_normalized}%
\end{equation}
where%
\begin{equation}
p\left(  d;\mu_{2},\sigma_{2},\mu_{2},\sigma_{2}\right)  \equiv p\left(
d;\sigma_{2}\right)  =\Phi\left(  \tfrac{d}{\sqrt{2}\sigma_{2}}\right)
-\Phi\left(  -\tfrac{d}{\sqrt{2}\sigma_{2}}\right)  =\operatorname{erf}\left(
\tfrac{d}{2\sigma_{2}}\right)  , \label{P_same}%
\end{equation}
so that $p_{0}=1$ if both individuals are of the same chronological age.

Setting the probability for people of different ages in relation to the
probability for people of the same age, as the normalized expression $\left(
\ref{P_normalized}\right)  $ does, provides a measure for a reasonable and
proportionate evaluation of the former. We will revisit it in section 4.1.2
within a more detailed discussion of the parameters.

\subsection{Individual case}

In order to compute the probability that one individual of chronological age
$\mu$ is within a certain mental age range $\left[  x_{1},x_{2}\right]  $ is%
\begin{equation}
p\left(  x_{1}\leq x\leq x_{2};\mu,\sigma\right)  =\int_{x_{1}}^{x_{2}%
}dxg\left(  x;\mu,\sigma^{2}\right)  =\Phi\left(  \tfrac{x_{2}-\mu}{\sigma
}\right)  -\Phi\left(  \tfrac{x_{1}-\mu}{\sigma}\right)  .
\end{equation}
In particular, this means for a symmetric span around $\mu$ and an upper and
lower bound, respectively,%
\begin{align}
p\left(  \mu-d\leq x\leq\mu+d;\mu,\sigma\right)   &  =\Phi\left(  \tfrac
{d}{\sigma}\right)  -\Phi\left(  -\tfrac{d}{\sigma}\right)
=\operatorname{erf}\left(  \tfrac{d}{\sqrt{2}\sigma}\right) \\
p\left(  x\geq x_{0};\mu,\sigma\right)   &  =1-\Phi\left(  \tfrac{x_{0}-\mu
}{\sigma}\right) \label{P_lower}\\
p\left(  x\leq x_{0};\mu,\sigma\right)   &  =\Phi\left(  \tfrac{x_{0}-\mu
}{\sigma}\right)  , \label{P_upper}%
\end{align}
where we used that $\underset{x\rightarrow-\infty}{\lim}\Phi\left(  x\right)
=0$ and $\underset{x\rightarrow\infty}{\lim}\Phi\left(  x\right)  =1$.

\section{Statistical expectation}

The above calculated probability is a mutual statement between two age groups,
that is, it states the likelihood of a selected pair of people of given ages
to be mentally compatible. Therefore, the computation of the statistical
expectation of compatible pairs is rather straightforward. However, the
statistical expectations within the groups themselves, like the number of
people in one group that have at least one or more compatible counterparts in
the other group, are interesting to consider as well; especially since there
might be overlaps, that is, multiple people could be compatible with the same
person or people.

Let us assume that one age group consists of $N_{1}$ and the other of $N_{2}$
individuals. The statistical expectation of the number of compatible pairs
$n_{12}$ is simply the total number of possible pairs multiplied by the
obtained probability:%
\begin{equation}
n_{12}=N_{1}N_{2}p. \label{Number_pairs}%
\end{equation}

The question is now, how many people of group 1 and group 2 compose these pairs.

It is quickly shown that the mean number of people in group 1 that are
mentally compatible with one randomly chosen individual in group 2 and vice
versa amount to%
\begin{equation}
\overline{n_{1\rightarrow2}}=N_{1}p\text{ and }\overline{n_{2\rightarrow1}%
}=N_{2}p. \label{Number_mean}%
\end{equation}

The probability that at least one member of group 1 is mentally compatible
with a randomly chosen member of group 2, amounts to $1-\left(  1-p\right)
^{N_{1}}$; so the expected value of members of group 2 that have at least one
mentally compatible counterpart in group 1 would be obtained by multiplying
the last result with $N_{2}$. For group 2, it works analogously. Hence the
expected values are%
\begin{equation}
n_{1}=N_{1}\left[  1-\left(  1-p\right)  ^{N_{2}}\right]  \text{ and }%
n_{2}=N_{2}\left[  1-\left(  1-p\right)  ^{N_{1}}\right]  .
\label{Number_individuals}%
\end{equation}

One could go even further and ask for the number of people in one group that
are mentally compatible with at least $k$ people in the other group. Since the
Binomials distribution,%
\begin{equation}
\Pr\left(  k;N,p\right)  =\tbinom{N}{k}p^{k}\left(  1-p\right)  ^{N-k},
\end{equation}
states the probability of exactly $k$ successes out of $N$ independent trials,
each subject to probability $p$, the probability of at least $k$ successes
ammounts to%
\begin{align}
\Pr\left(  m\geq k;N,p\right)   &  =1-%
%TCIMACRO{\tsum \limits_{m=0}^{k-1}}%
%BeginExpansion
{\textstyle\sum\limits_{m=0}^{k-1}}
%EndExpansion
\Pr\left(  m;N,p\right)  =1-%
%TCIMACRO{\tsum \limits_{m=0}^{k-1}}%
%BeginExpansion
{\textstyle\sum\limits_{m=0}^{k-1}}
%EndExpansion
\tbinom{N}{m}p^{m}\left(  1-p\right)  ^{N-m}\nonumber\\
&  \approx1-\tfrac{1}{\sqrt{2\pi Np\left(  1-p\right)  }}\int_{0}%
^{k}dxe^{-\frac{1}{2}\frac{\left(  x-Np\right)  ^{2}}{Np\left(  1-p\right)  }%
}\label{Probability_individuals_k}\\
&  =1-\Phi\left(  \tfrac{k-Np}{\sqrt{Np\left(  1-p\right)  }}\right)
+\Phi\left(  -\sqrt{\tfrac{Np}{1-p}}\right)  ,\nonumber
\end{align}
where the approximation by a normal distribution is justified for sufficiently
big $N$ fulfilling the condition%
\begin{equation}
N>9\max\left(  \tfrac{p}{1-p},\tfrac{1-p}{p}\right)  .
\end{equation}

Note: $\left(  \ref{Number_individuals}\right)  $ can be inferred from setting
$k=1$ in the first line of $\left(  \ref{Probability_individuals_k}\right)  $.

\section{Parameter discussion}

In this section, we infer ranges for the standard deviation $\sigma$ and the
mental age difference $d$ from empirical data and a detailed analysis of the
pdf, followed by a discussion of some interesting implications.

\subsection{Ranges of parameters}

\subsubsection{Standard deviation}

From the findings in
\cite{SchutteMalouffHall,TapiaMarshII,Bar-On,EQYoungAdults,EQOlderChildren,Amantha}%
, some of which are distinguishing between multiple sub-scales regarding
different components of emotional intelligence, we obtain a reliable span of
values for the statistical dispersions of the most significant domains of
mental age, ranging over various age groups.\ One observation from the studies
is that the average scores from the applied tests are about proportional to
the chronological ages of the participating groups. This suggests that results
scattered around the mean of one age group likely evolve accordingly as age
progresses -- and so the dispersion of those results. Hence one may write%
\begin{equation}
s\equiv\tfrac{\sigma}{\mu}=const. \label{sigma-mu ratio}%
\end{equation}

Regardless of the considered component of (emotional) intelligence, standard
deviations obtained from said works are mostly scattered in between 10\% and
slightly above 20\% of the mean value\footnote{Bar-On and others used to
normalize their score histograms to the common IQ-distribution with $\mu=100$
and $\sigma=15$.}. The full range of the scatter presumably results in part
from focusing on different groups of people and factors of interest throughout
the studies, like culture, gender etc.

In the following calculations, we will account for the bounds:%
\begin{equation}
0.1\mu\leq\sigma\leq0.2\mu. \label{Standard Deviation}%
\end{equation}

We will see that the outcome could depend quite sensitively on slight shifts
in that percentage, which is why a separate computation for the interval
bounds is definitely sensible.

\subsubsection{Mental age difference}

The allowed mental age difference is, as already mentioned, subject to
individual and cultural morality; however, since it would not really make
sense to choose the allowed mental age difference stricter than the natural
mental dispersion within one age group, it is reasonable to take the first
standard deviation of the younger of two considered age groups as its lower
bound:%
\begin{equation}
d\geq\min\left(  \sigma_{1},\sigma_{2}\right)  . \label{Mental Age Difference}%
\end{equation}

There is also another interesting and meaningful way of deriving a measure for
$d$. Instead of looking at the pdf for mental age itself, as done so far, one
may ask for the pdf of mental age differences within the same chronological
age group. This can be obtained in an analogous fashion to above from the
convolution of the pdf for mental age with itself:%
\begin{align}
h\left(  d\right)   &  =\int_{-\infty}^{\infty}dxg\left(  x;\mu,\sigma
^{2}\right)  g\left(  x-d;\mu,\sigma^{2}\right) \nonumber\\
&  =\int_{-\infty}^{\infty}dxg\left(  x;\mu,\sigma^{2}\right)  g\left(
d-x;-\mu,\sigma^{2}\right) \\
&  =\left(  g\ast g\right)  \left(  d;0,2\sigma^{2}\right)  ,\nonumber
\end{align}
from which we read off%
\begin{equation}
\sigma_{d}=\sqrt{2}\sigma.
\end{equation}

Note that, because $d\geq0$, the true pdf of mental age differences within the
same chronological age group is a half-normal distribution:%
\begin{equation}
\hat{h}\left(  d\right)  =2h\left(  d\right)  \text{ for }d\geq0.
\end{equation}
The sought measure is the mean mental age difference, so the expectation of
$d$ under this pdf:%
\begin{equation}
\left\langle d\right\rangle _{\hat{h}}=\sqrt{\tfrac{2}{\pi}}\sigma_{d}%
=\tfrac{2}{\sqrt{\pi}}\sigma\approx1.128\,\sigma,
\label{Mental Age Difference 2}%
\end{equation}
which is going to be the main reference value for $d$ henceforth.

The dispersion of values around this mean amounts to%
\begin{equation}
\hat{\sigma}_{d}=\sqrt{\left\langle \left(  d-\left\langle d\right\rangle
_{\hat{h}}\right)  ^{2}\right\rangle _{\hat{h}}}=\sigma_{d}\sqrt{1-\tfrac
{2}{\pi}}=\sigma\sqrt{2\left(  1-\tfrac{2}{\pi}\right)  }\approx0.853\sigma.
\label{Mental Age Dispersion}%
\end{equation}

The scope for the mental age differences can now be inferred from the
intersection of the interval given by $\left\langle d\right\rangle _{\hat{h}%
}\pm\hat{\sigma}_{d}$ and the lower bound $d\geq\sigma$ in virtue of $\left(
\ref{Mental Age Difference}\right)  $; hence, with $\left(
\ref{Mental Age Difference 2}\right)  -\left(  \ref{Mental Age Dispersion}%
\right)  $ and the fact that $\left\langle d\right\rangle _{\hat{h}}%
-\hat{\sigma}_{d}\approx0.276\sigma<\sigma$, we have%
\begin{equation}
\sigma\leq d\leq1.981\sigma. \label{Mental Age Scope}%
\end{equation}

For different chronological ages, $\mu_{1}\neq\mu_{2}$, it is thereby, in
accordance with $\left(  \ref{Mental Age Difference}\right)  $,%
\begin{equation}
\sigma=\min\left(  \sigma_{1},\sigma_{2}\right)  .
\end{equation}

Another interesting consequence are the actual probabilities for two randomly
chosen people from the same chronological age group. Using $\left(
\ref{P_same}\right)  $, we find%
\begin{equation}
p\left(  d=t\sigma,\mu,\sigma,\mu,\sigma\right)  \equiv p\left(  t\right)
=\Phi\left(  \tfrac{t}{\sqrt{2}}\right)  -\Phi\left(  -\tfrac{t}{\sqrt{2}%
}\right)  =\operatorname{erf}\left(  \tfrac{t}{2}\right)  . \label{P_same_2}%
\end{equation}

Note that this result is independent of both chronological age and standard
deviation, since substituting $d$ with any constant multiple of $\sigma$
cancels the contribution of $\sigma$ completely. Tab. 1 lists the same-age
probabilities of a few special cases for $d$:%
\begin{align*}
&
\begin{tabular}
[c]{|l|l|l|l|l|}\hline
$d$ & $\sigma$ & $\tfrac{2}{\sqrt{\pi}}\sigma$ & $\sqrt{2}\sigma$ &
$\left\langle d\right\rangle _{\hat{h}}+\hat{\sigma}_{d}$\\\hline
$p$ & $0.52$ & $0.58$ & $0.68$ & $0.84$\\\hline
\end{tabular}
\\
&  \text{Tab 1}\colon\text{ }p\text{ vs. }d\text{ for some special cases}%
\end{align*}

So, the probability that two randomly chosen people from the same
chronological age group are mentally compatible is less than 60\%, if the
benchmark $\left(  \ref{Mental Age Difference 2}\right)  $ is applied. This is
an essential finding, as it provides a frame of reference for interpreting
probabilities regarding different chronological age groups in the right
(moral/ethical/legal) proportion.

\subsection{Age ratio}

Assuming $\mu_{1}\geq\mu_{2}$ and writing, as in the last subsection,
$\sigma_{1,2}=s_{1,2}\mu_{\,1,2}$ and $d=t\sigma_{2}$, we find%
\begin{equation}
p=\Phi\left(  \tfrac{\frac{\mu_{1}}{\mu_{2}}-1+s_{2}t}{\sqrt{s_{1}^{2}\left(
\frac{\mu_{1}}{\mu_{2}}\right)  ^{2}+s_{2}^{2}}}\right)  -\Phi\left(
\tfrac{\frac{\mu_{1}}{\mu_{2}}-1-s_{2}t}{\sqrt{s_{1}^{2}\left(  \frac{\mu_{1}%
}{\mu_{2}}\right)  ^{2}+s_{2}^{2}}}\right)  .
\end{equation}
So, besides the coefficients $s_{1}$, $s_{2}$, and $t$, the probability only
depends on the ratio of the chronological ages -- not their absolute values.
If the relationship $\left(  \ref{sigma-mu ratio}\right)  $ was not applied,
then the absolute values of $\mu_{1}$ and $\mu_{2}$ might be implicit in the
choices of $s_{1}$ and $s_{2}$; unless we assume $s_{1}=s_{2}$, in which case
the latter, including their potential age-dependences, cancel.

One interesting implication is that, for example, the probability of mental
compatibility between two people aged 16 and 24 is the same as for two people
aged 24 and 36.

Important note: This result is to be interpreted under the consideration that
the same factor $t$ is chosen for the allowed mental age difference; in
particular, a randomly chosen 16 year-old is as likely to be mentally at most
1.6 years apart from a randomly chosen 24 year-old as a randomly chosen 24
year-old is to be mentally at most 2.4 years apart from a randomly chosen 36
year-old, if one chooses the lowest bound with $s_{1}=s_{2}=0.1$ and $t=1$.

\section{Applications}

We apply the findings derived above to real-life examples. The implications of
the math derived above are illustrated by computing statistical expectations
for the mental compatibility of particular groups of individuals and analyzing
popular age rules.

\subsection{High school and college students}

The usual age range in high school is 14-18 years and in college 18-22 years.
If we take the youngest age of 14 years as a reference, then the probability
that an individual in high school or college is mentally compatible\ with
someone of chronological age 14 is illustrated in fig. 1. It is very
interesting to see that about one ninth of people aged 14 and 18 are mentally
compatible under minimum conditions; especially if contrasted with the fact
that the probability for two 14 year-olds under the same conditions is only a
bit more than one half.%
\begin{align*}
&
%TCIMACRO{\FRAME{itbpF}{7.1269in}{3.0805in}{0in}{}{}{figure1.eps}%
%width 7.1269in;  height 3.0805in;  depth 0in;  original-width 10.1408in;
%original-height 4.3613in;  cropleft "0";  croptop "1";  cropright "1";
%cropbottom "0";  filename 'figure1.eps';file-properties "XNPEU";}} }%
%BeginExpansion
{\includegraphics[
height=3.0805in,
width=7.1269in
]%
{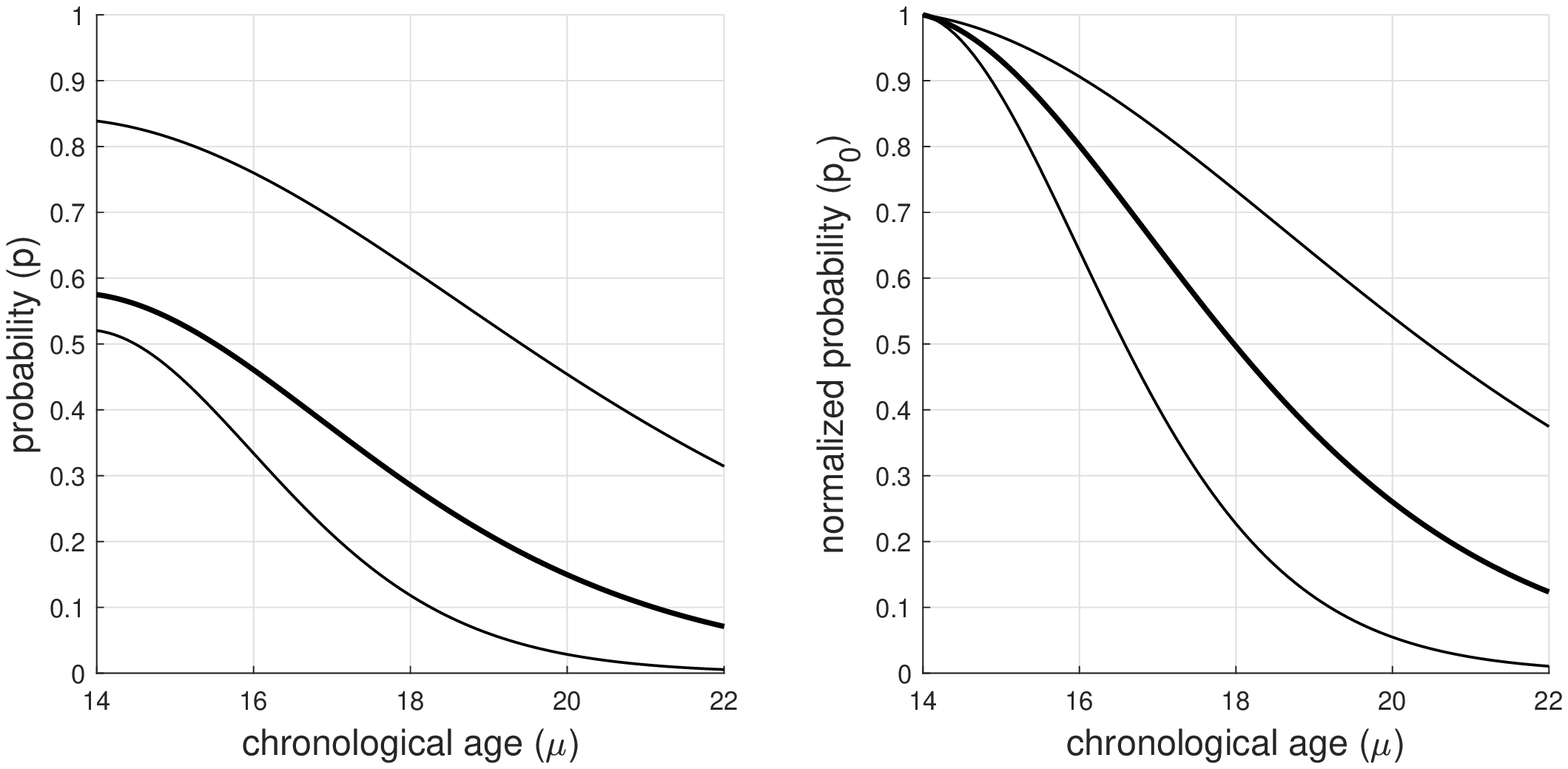}%
}
%EndExpansion
\\
&
\begin{tabular}
[c]{l}%
Fig. 1: Mental compatibility between age cohorts in high school/college and
14-y.o.'s; thin lines\\
represent the upper and lower bounds and thicker lines the benchmark for
$s=0.15$ and $t=\frac{2}{\sqrt{\pi}}$.\\
\textit{Left: }regular probabilities; \textit{right:} normalized
probabilities.
\end{tabular}
\end{align*}

Another example: Considering the scopes of $\sigma$ and $d$, as explained in
subsection 4.1, two people of chronological ages 16 and 20, respectively, are
mentally compatible within a probability range $0.16\leq p\leq0.65$; the
benchmark at $s=0.15$ and $t=\frac{2}{\sqrt{\pi}}$ is $p=0.33$. So at least
about one sixth of pairs formed by 16 and 20 year-olds can be considered
mentally compatible within the present framework.

These results are significant, since they strongly suggest the potential of
collaboration between students of different ages and merged classes of
different grades for special projects -- possibly even between high school and
college students.

According to the National Center for Education Statistics (NCES) \cite{NCES},
the number of students attending grades 9 through 12 at public high schools in
the United States was about 15.1 Million in Fall 2017. Assuming an approximate
equipartition of this number across the four grades, every age group
encompassed about 3.78 Million students. In the following, we will compute a
few statistical expected values, as outlined in section 3, taking the minimum
case calculated above with $p\approx\frac{1}{9}$:

\begin{itemize}
\item With $\left(  \ref{Number_pairs}\right)  $ at least $\left(
3.78\times10^{6}\right)  ^{2}\cdot\frac{1}{9}\approx1.59\times10^{12}=1.59$
trillion (out of $\left(  3.78\times10^{6}\right)  ^{2}\approx14.3$ trillion
possible) pairs of American public high school students aged 14 and 18,
respectively, were mentally compatible in Fall 2017. (Note: The vast size of
the last result is due to a huge number of intersections among the pairs)

\item The expected number of students aged 18 who are mentally compatible with
a randomly chosen 14 year-old\ -- and vice versa -- amounts with $\left(
\ref{Number_mean}\right)  $\ to at least $3.78\times10^{6}\cdot\frac{1}%
{9}=\allowbreak420,000$.

\item According to $\left(  \ref{Number_individuals}\right)  $, the expected
values within both age groups, counting how many students have one or more
mentally compatible counterparts in the other group, are infinitesimally close
to $3.78$ Million; that is, nearly every 14 year-old and nearly every 18
year-old has one or more -- based on the average of $420$ thousand, rather a
great many of -- persons in the other group whom they are mentally compatible with.
\end{itemize}

\subsection{Age limits}

If we were to set up an age limit in a way that is consistent with the present
mathematical framework, we would first of all define a mental -- not
chronological -- age limit $x_{\min}$, as most subjects pertaining to age
laws, like adulthood, really mean a state of mind rather than cell aging
(albeit by correlation -- not by causation). Then we would need to agree on a
limit (minimum or maximum) probability/expected relative frequency $p_{\lim}$
to decide on the threshold share of people within a \textquotedblleft
qualifying\textquotedblright\ chronological age group that are above or below
the stipulated mental age, where $\mu_{\lim}$, with associated standard
deviation $\sigma_{\lim}=s\mu_{\lim}$, is the limit chronological age
fulfilling that condition.

As a consequence, $p_{\lim}=0.5$ generally implies $x_{\lim}=\mu_{\lim}$,
which also follows immediately from the axial symmetry of a Gaussian around
the vertical axis at its mean value, in case of negligible skewness. This
means in particular for $p_{\lim}\gtrless0.5$: $\mu_{\min}\gtrless x_{\min}$
and $\mu_{\max}\lessgtr x_{\max}$.

For lower age limits ($\lim=\min$), it is, together with $\left(
\ref{P_lower}\right)  $,%
\begin{equation}
p_{\min}=p\left(  x\geq x_{\min};\mu_{\min},\sigma_{\min}\right)
=1-\Phi\left(  \tfrac{x_{\min}-\mu_{\min}}{s\mu_{\min}}\right)  ,
\end{equation}
which has to be solved for $\mu_{\min}$ to get the minimum chronological age
for the sought age limit,%
\begin{equation}
\mu_{\min}=\tfrac{x_{\min}}{1+s\Phi^{-1}\left(  1-p_{\min}\right)  },
\label{Age_min}%
\end{equation}
where $\Phi^{-1}$ is the inverse cdf or the quantile function.

Analogously, for upper age limits ($\lim=\max$), we find with $\left(
\ref{P_upper}\right)  $,%
\begin{equation}
p_{\max}=p\left(  x\leq x_{\max};\mu_{\max},\sigma_{\max}\right)  =\Phi\left(
\tfrac{x_{\max}-\mu_{\max}}{s\mu_{\max}}\right)  ,
\end{equation}
and hence%
\begin{equation}
\mu_{\max}=\tfrac{x_{\max}}{1+s\Phi^{-1}\left(  p_{\max}\right)  }.
\label{Age_max}%
\end{equation}

\subsubsection{Example 1: age of majority}

The age of majority is the age at which human beings enter adulthood. Although
the complexity behind the complete definition is not relevant for our puposes
here, it shall be remarked that it has varied from culture to culture
throughout history and so given rise to a multiplicity of age laws in the
world \cite{Hamilton}. In the vast majority of countries, that number was set
at 18, in some it is as small as 14, and in some as big as 21.

In reality, a limit for the chronological -- not mental -- age is taken as a
measure; set mostly to $\mu_{\min}=18$, the expression $\left(  \ref{Age_min}%
\right)  $ implies the corresponding mental age limits, depending on the
choices of $s$ and $p_{\min}$, as depicted in fig. 2 below. It is obvious that
those choices sensitively affect the mental age bar, which raises the question
of the underlying reasoning.

\emph{Upshot:} The fact that the legislators behind such age laws only refer
to the chronological age either means that they don't ascribe any relevance to
the concept of mental age per se or simply equate both ages. The latter case
would imply the choice $p_{\min}=0.5$, for which $x_{\min}=\mu_{\min}$ for all
values of $s$, if one presumed reasoning only remotely akin to the present
work. Since that is rather unlikely, however, it seems that at best there is a
mere focus on the statistical average disregarding the significant
implications of the standard deviation, as detailed in this work; or they are
well aware of it but dodge the question for lack of adequate theoretical support.

\subsubsection{Example 2: senior age}

The age from which a person is defined as \textquotedblleft
elder\textquotedblright\ ranges between 60 and 65 years. If one defines
\textquotedblleft senior\textquotedblright\ as a mostly mental state, one may
analyze it as an upper age limit in an analogous fashion to the previous example.

Since those age numbers are in reality also solely of chronological nature, we
would use $\left(  \ref{Age_max}\right)  $ to find the corresponding mental
age limits within the range of parameters. Fig. 2 depicts them for $\mu_{\max
}=60$.

The upshot regarding that social system of chronological age limits would be
similar to the one stated above for the age of majority.%
\[%
%TCIMACRO{\FRAME{itbpFU}{7.0491in}{3.0805in}{0in}{\Qcb{Fig. 2$\colon$ Mental
%age limits vs. limit probabilities for given chronological age limits: 18
%(l.), 60 (r.).}}{}{figure2.eps}{\special{ language "Scientific Word";
%type "GRAPHIC";  maintain-aspect-ratio TRUE;  display "USEDEF";
%valid_file "F";  width 7.0491in;  height 3.0805in;  depth 0in;
%original-width 10.4209in;  original-height 4.5312in;  cropleft "0";
%filename 'figure2.eps';file-properties "XNPEU";}} }%
%BeginExpansion
{\parbox[b]{7.0491in}{\begin{center}
\includegraphics[
height=3.0805in,
width=7.0491in
]%
{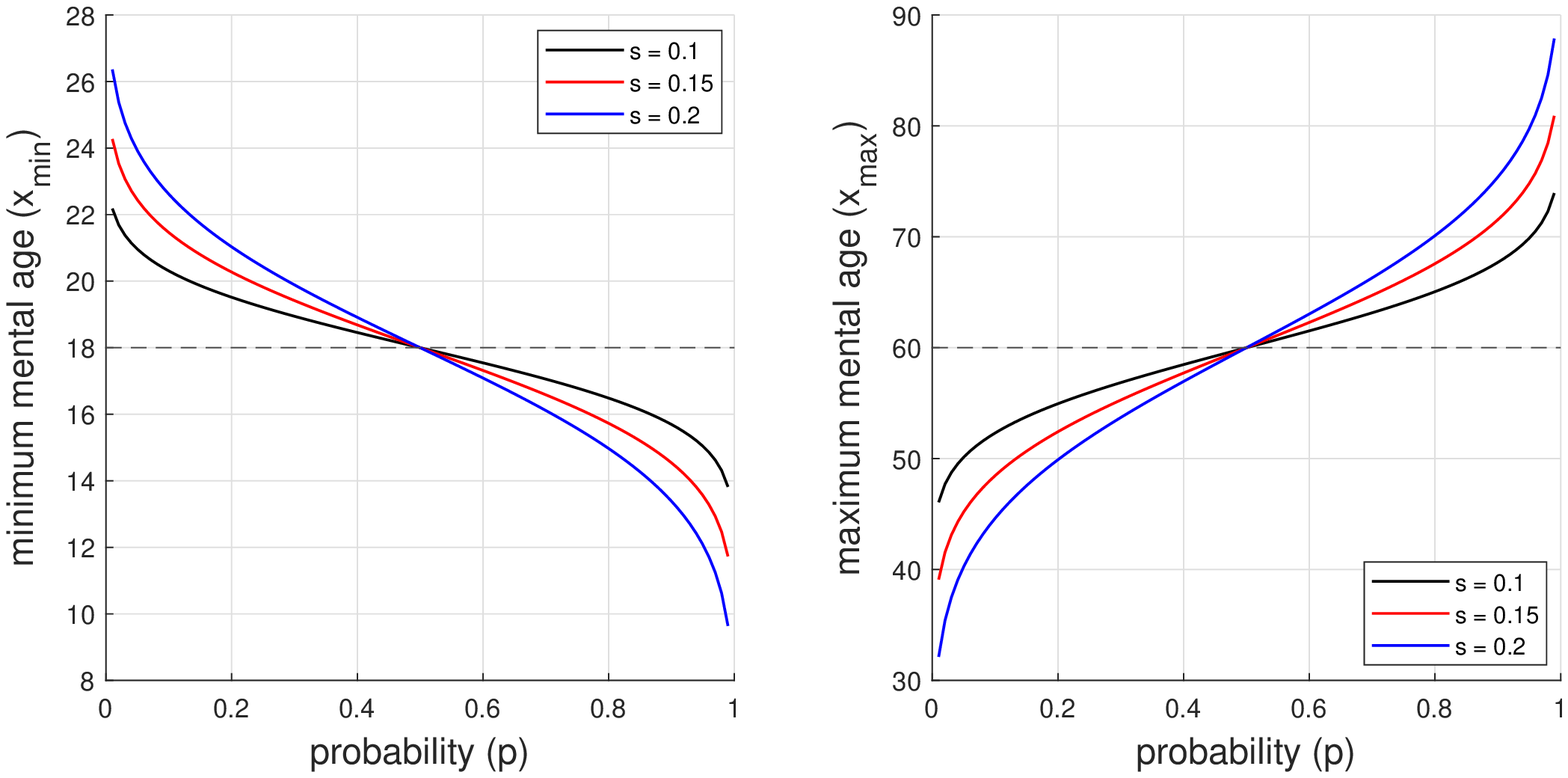}%
\\
Fig. 2$\colon$ Mental age limits vs. limit probabilities for given
chronological age limits: 18 (l.), 60 (r.).
\end{center}}}
%EndExpansion
\]

\subsection{The \textquotedblleft half-your-age-plus-seven\textquotedblright%
\ rule}

A fairly common rule of thumb in determining a lower (chronological) age bound
to a \textquotedblleft socially acceptable\textquotedblright\ gamut with
regard to dating relationships is the so-called \textquotedblleft
half-your-age-plus-seven\textquotedblright\ rule \cite{DiDonato}, which is
supposed to be taken literally:%
\begin{equation}
\mu_{\min}=\tfrac{1}{2}\mu+7, \label{Age_Minimum}%
\end{equation}
where $\mu$ is the chronological age of the inquiring person.

This rule can also be inverted to infer an upper age bound:%
\begin{equation}
\mu_{\max}=2\mu-14. \label{Age_Maximum}%
\end{equation}

Albeit of unclear origin, this rule is reported as once not being meant as a
boundary for appropriateness but rather a measure for the ideal (younger) age
of a man's bride \cite{LockerLampson,O'Rell,Herbert}. In any case, it seems to
be quite commonly applied these days for the above-mentioned reason, so that
it is definitely worth a closer inspection in the framework of the math
presented here.

In view of mental age compatibility, finding a bound for chronological ages
would most sensibly connect to defining a bound for the probability $\left(
\ref{P_Result}\right)  $. For the remainder of this subsection, $\mu$ is the
younger age and $\Delta$ the maximum difference upward in terms of
\textquotedblleft social acceptibility\textquotedblright; then we would write,
after some algebra,%

\begin{equation}
p_{\min}=\Phi\left(  \tfrac{\frac{\Delta}{\mu}+ts_{1}}{\sqrt{s_{1}^{2}%
+s_{2}^{2}\left(  1+\frac{\Delta}{\mu}\right)  ^{2}}}\right)  -\Phi\left(
\tfrac{\frac{\Delta}{\mu}-ts_{1}}{\sqrt{s_{1}^{2}+s_{2}^{2}\left(
1+\frac{\Delta}{\mu}\right)  ^{2}}}\right)  ,
\label{Implicit relation_minimum}%
\end{equation}
where we used $\sigma_{1}=s_{1}\mu$, $\sigma_{2}=s_{2}\left(  \mu
+\Delta\right)  $, and $d=t\sigma_{1}$ with $t\geq1$ in accordance with
$\left(  \ref{Mental Age Scope}\right)  $.

The last expression is equivalent to the relation%
\begin{equation}
\tfrac{\Delta}{\mu}=f\left(  p_{\min,}s_{1},s_{2},t\right)  ,
\end{equation}
where the RHS is a constant, unless the parameters $s_{1}$ and $s_{2}$ are
assumed to harbor a dependence on $\mu$ and $\Delta$. This is congruent with
subsection 4.2.

Assuming $s_{1},s_{2}=const.$ in accordance with $\left(  \ref{sigma-mu ratio}%
\right)  $, $\left(  \ref{Implicit relation_minimum}\right)  $ requires a
proportional relationship,%
\begin{equation}
\Delta=m\mu\text{ or }\mu_{\max}=\left(  m+1\right)  \mu, \label{Delta_mu}%
\end{equation}
in order for the result to be a constant; otherwise, the relation would have a
more complicated form, although it really appears that the empirically
determined values of $s$ are rather dense within the interval $\left[
0.1,0.2\right]  $ and thus have not much of an impact within their scope of variation.

Rewriting $\left(  \ref{Age_Maximum}\right)  $ to $\tfrac{\Delta}{\mu
}=1-\tfrac{14}{\mu}$, one sees directly that $\Delta$ and $\mu$ deviate
significantly from being proportionial for realistic values of $\mu$, since
the approximation $\frac{\Delta}{\mu}\approx1$ would require $\mu\geq280$ to
have an error $\tfrac{14}{\mu}\leq0.05$ (or $\mu\geq140$ if one goes with
$\left(  \ref{Delta_mu}b\right)  $). Even allowing $s_{1}$ and $s_{2}$ to vary
within $\left[  0.1,0.2\right]  $, $\left(  \ref{Implicit relation_minimum}%
\right)  $ cannot nearly be fulfilled with the substitution $\tfrac{\Delta
}{\mu}=1-\tfrac{14}{\mu}$ (see fig. 3).%

\begin{align*}
&
%TCIMACRO{\FRAME{itbpF}{3.5855in}{2.693in}{0in}{}{}{figure3.eps}%
%width 3.5855in;  height 2.693in;  depth 0in;  original-width 6.0614in;
%original-height 4.5411in;  cropleft "0";  croptop "1";  cropright "1";
%cropbottom "0";  filename 'figure3.eps';file-properties "XNPEUR";}} }%
%BeginExpansion
{\includegraphics[
height=2.693in,
width=3.5855in
]%
{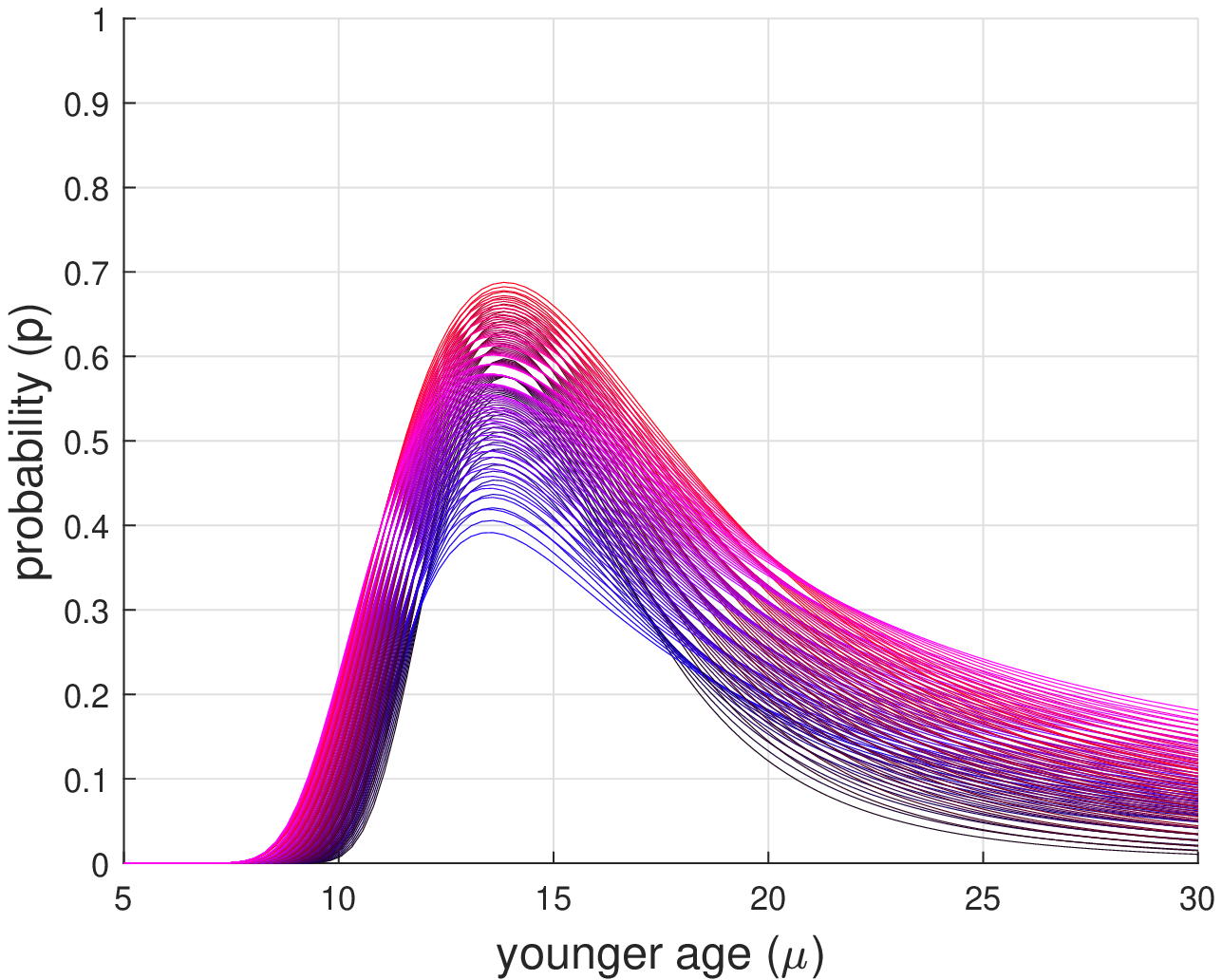}%
}
%EndExpansion
\\
&
\begin{tabular}
[c]{l}%
Fig. 3: The half-your-age-plus-seven rule applied\\
to the probability formalism with various combi-\\
nations of $s_{1},s_{2}\in\left[  0.1,0.2\right]  $%
\end{tabular}
\end{align*}

\emph{Upshot:} If one sets the \textquotedblleft
half-your-age-plus-seven\textquotedblright\ rule\ as a basis, one does not
arrive at a consistent lower-bound statement for the probability, as far as
the current mathematical framework is concerned. Briefly, the
\textquotedblleft half-your-age-plus-seven\textquotedblright\ rule\ is not
compatible with the present findings. Given that the formalism presented here
is a consistently established mathematical model based on empirical data from
scientific studies and the \textquotedblleft
half-your-age-plus-seven\textquotedblright\ rule\ a loose rule of thumb
without scientific foundation nor clear origin reflecting the purpose it is
claimed to have these days, the former merits preferential consideration by
any rational means.

We conclude this subsection with a table displaying the values for $m$ from
the proportional relationship between $\Delta$ and $\mu$, according to
$\left(  \ref{Delta_mu}\right)  $, for a few special cases of the $s$-values
(we assume $s_{1}=s_{2}$ independent of age) and $p_{\min}$:%
\begin{align*}
&
\begin{tabular}
[c]{|l|l|l|l|}\hline
& $p_{\min}=0.05$ & $p_{\min}=0.1$ & $p_{\min}=0.15$\\\hline
$s=0.1$ & $0.39$ & $0.32$ & $0.28$\\\hline
$s=0.15$ & $0.64$ & $0.51$ & $0.43$\\\hline
$s=0.2$ & $0.92$ & $0.71$ & $0.59$\\\hline
\end{tabular}
\\
&  \text{Tab. 2}\colon\text{ }m\text{-values for }p_{\min}\text{ vs. }s
\end{align*}

\section{Conclusion}

We have set up a formula for computing the probability that two individuals of
given chronological ages are within a certain mental age difference. From
there, we derived expressions for statistical expectations and condensed
ranges and benchmarks for the parameters through empirical data from studies
and in-depth analysis of the presented mathematical structure. With these
findings, we could investigate the validity of popular assumptions about age
differences and existing rules regarding age limits through detailed
qualitative and quantitative elaboration.

For the entirety of high school and college students in the US from a given
year, we computed the probabilities between cohorts that are further apart in
age and derived the statistical expectations of both pairs and individuals
that are mentally compatible. The results showed clearly that even between
14-year-olds and 18-year-olds there is a noteable number of mentally
compatible people that is far from any negligence.

We critically examined the sensibility of socially imposed chronological age
limits by depicting the noteable span of associated mental ages following from
the presented formalism within the scope of parameters and concluded a lack of
the concept of mental age in such legislations.

Finally, we questioned the \textquotedblleft
half-your-age-plus-seven\textquotedblright\ rule and found that it is not
compatible with the presented statistical formalism. Given that a both
empirically supported and mathematically consistent fundament is in clear
contrast to a loose rule of thumb of unclear origin, said rule could be shown
to stand on not too solid ground.

Altogether, we derived a formal basis on which we could bring forward
substantial reasoning to question the largely prevailing construct of ideas
around age.

More in-depth directed research in the field, i.a. in form of more
specifically targeted studies, will help refine the methodology, particularly
in terms of parameter accuracy, and increase the precision -- and hence
expressiveness -- of results.

\bigskip

\noindent\textbf{Acknowledgments}

First and foremost, I want to thank my mother, Karla, for her loving support
during which this work emerged. I also would like to thank Werner D\"{a}ppen
for a stimulating exchange of minds and Walter Unglaub for helpful assistance
with technical issues.

\appendix

\section{Appendix: Error discussion}

\subsection{Outer sectors}

As mentioned earlier, the outer age sectors ($x<0$), which have been included
in the integral to arrive at the closed-form solution $\left(  \ref{P_Result}%
\right)  $, have negligible contribution. This can be seen by secting $\left(
\ref{P_Integral}\right)  $, which is the product of two Gaussians integrated
over the area spanned by $\left\{  \left(  x_{1},x_{2}\right)  \in%
%TCIMACRO{\U{211d} }%
%BeginExpansion
\mathbb{R}
%EndExpansion
^{2}|x_{1}-d\leq x_{2}\leq x_{1}+d\right\}  $. In particular, one decomposes
this domain into lines, defined by $x_{2}=x_{1}+\Delta$ ($-d\leq\Delta\leq
d$), and looks at the corresponding one-dimensional slices of the product
function,
\begin{equation}
\tfrac{1}{2\pi\sigma_{1}\sigma_{2}}e^{-\frac{1}{2}\left[  \left(  \frac
{x_{1}-\mu_{1}}{\sigma_{1}}\right)  ^{2}+\left(  \frac{x_{1}+\Delta-\mu_{2}%
}{\sigma_{2}}\right)  ^{2}\right]  }=\tfrac{1}{2\pi\sigma_{1}\sigma_{2}%
}e^{-\frac{1}{2}\frac{\left(  \mu_{1}-\mu_{2}+\Delta\right)  ^{2}}{\sigma
_{1}^{2}+\sigma_{2}^{2}}}e^{-\frac{1}{2}\left(  \frac{x_{1}-\mu_{12}}%
{\sigma_{12}}\right)  ^{2}}, \label{Product Function_Slice}%
\end{equation}
where%
\begin{equation}
\mu_{12}=\tfrac{\left(  \mu_{2}-\Delta\right)  \sigma_{1}^{2}+\mu_{1}%
\sigma_{2}^{2}}{\sigma_{1}^{2}+\sigma_{2}^{2}}\text{ and }\sigma_{12}%
=\tfrac{\sigma_{1}\sigma_{2}}{\sqrt{\sigma_{1}^{2}+\sigma_{2}^{2}}}.
\end{equation}
The trick is now to compare the integral of $\left(
\ref{Product Function_Slice}\right)  $ over the outer sector, $\left]
-\infty,0\right]  $, with the entire one used above, $\left]  -\infty
,\infty\right[  $, for which the rewritten form on the RHS is very helpful:%
\[
\frac{\int_{-\infty}^{0}dx_{1}e^{-\frac{1}{2}\left(  \frac{x_{1}-\mu_{12}%
}{\sigma_{12}}\right)  ^{2}}}{\int_{-\infty}^{\infty}dx_{1}e^{-\frac{1}%
{2}\left(  \frac{x_{1}-\mu_{12}}{\sigma_{12}}\right)  ^{2}}}=\tfrac{1}%
{\sqrt{2\pi}\sigma_{12}}\int_{-\infty}^{0}e^{-\frac{1}{2}\left(  \frac
{x-\mu_{12}}{\sigma_{12}}\right)  ^{2}}=\Phi\left(  -\tfrac{\mu_{12}}%
{\sigma_{12}}\right)  .
\]
The argument on the very right can be estimated like%
\[
-\tfrac{\mu_{12}}{\sigma_{12}}=-\tfrac{\left(  \mu_{2}-\Delta\right)
\sigma_{1}^{2}+\mu_{1}\sigma_{2}^{2}}{\sigma_{1}\sigma_{2}\sqrt{\sigma_{1}%
^{2}+\sigma_{2}^{2}}}\leq-\tfrac{\mu_{2}\sigma_{1}^{2}+\mu_{1}\sigma_{2}^{2}%
}{\sigma_{1}\sigma_{2}\sqrt{\sigma_{1}^{2}+\sigma_{2}^{2}}}+\tfrac{\sigma_{1}%
}{\sigma_{2}\sqrt{\sigma_{1}^{2}+\sigma_{2}^{2}}}d<-\tfrac{\mu_{2}-d}%
{\sigma_{2}},
\]
where we assumed $\frac{\mu_{1}}{\sigma_{1}}=\frac{\mu_{2}}{\sigma_{2}}$, as
explained in section 4. Furthermore, assuming $0.1\leq\frac{\sigma}{\mu}%
\leq0.2$ and $d\sim\sigma$, we have%
\[
\Phi\left(  -\tfrac{\mu_{12}}{\sigma_{12}}\right)  <\Phi\left(  -\tfrac
{\mu_{2}-d}{\sigma_{2}}\right)  \leq\Phi\left(  -4\right)  \approx
3.17\times10^{-5}.
\]

Since this ratio bound holds for every slice of the integration domain, the
contribution from the sector $x<0$ is negligible. The same applies to the
sector $x\gg\max\left(  \mu_{1},\mu_{2}\right)  $ far beyond human age.

\subsection{Error propagation}

The main uncertainty in this endeavor comes from the proper estimation of the
standard deviations, $\sigma_{1}$ and $\sigma_{2}$, and the choice of the
allowed mental age difference, $d$. The error propagation resulting from these
amounts to%
\begin{align}
\Delta p\left(  d;\mu_{1},\sigma_{1},\mu_{2},\sigma_{2}\right)   &
=\tfrac{\Delta d}{\sqrt{2\pi\left(  \sigma_{1}^{2}+\sigma_{2}^{2}\right)  }%
}\left[  e^{-\frac{1}{2}\frac{\left(  \mu_{1}-\mu_{2}+d\right)  ^{2}}%
{\sigma_{1}^{2}+\sigma_{2}^{2}}}+e^{-\frac{1}{2}\frac{\left(  \mu_{1}-\mu
_{2}-d\right)  ^{2}}{\sigma_{1}^{2}+\sigma_{2}^{2}}}\right]
\label{Error Propagation}\\
&  +\tfrac{\sigma_{1}\Delta\sigma_{1}+\sigma_{2}\Delta\sigma_{2}}{\sqrt{2\pi
}\left(  \sigma_{1}^{2}+\sigma_{2}^{2}\right)  ^{\frac{3}{2}}}\left\vert
\left(  \mu_{1}-\mu_{2}+d\right)  e^{-\frac{1}{2}\frac{\left(  \mu_{1}-\mu
_{2}+d\right)  ^{2}}{\sigma_{1}^{2}+\sigma_{2}^{2}}}-\left(  \mu_{1}-\mu
_{2}-d\right)  e^{-\frac{1}{2}\frac{\left(  \mu_{1}-\mu_{2}-d\right)  ^{2}%
}{\sigma_{1}^{2}+\sigma_{2}^{2}}}\right\vert .\nonumber
\end{align}

Let us compare the impacts of the single errors. The ratio of the standard
deviation and the mental age part is%
\begin{equation}
\tfrac{\Delta p|_{\Delta d=0}}{\Delta p|_{\Delta\sigma_{1,2}=0}}=\tfrac
{1}{\sigma_{1}^{2}+\sigma_{2}^{2}}\left\vert \left(  \mu_{1}-\mu_{2}\right)
\tanh\left[  \tfrac{2d\left(  \mu_{1}-\mu_{2}\right)  }{\sigma_{1}^{2}%
+\sigma_{2}^{2}}\right]  +d\right\vert \tfrac{\sigma_{1}\Delta\sigma
_{1}+\sigma_{2}\Delta\sigma_{2}}{\Delta d}. \label{Error Ratio}%
\end{equation}

If we take $d=a\sigma_{1}$, $\mu_{1}-\mu_{2}=b\sigma_{1}$, and $\sigma
_{1}=c\sigma_{2}$, with $a\,,b,c>1$, this becomes%
\begin{equation}
\tfrac{\Delta p|_{\Delta d=0}}{\Delta p|_{\Delta\sigma_{1,2}=0}}=\tfrac
{c}{1+c^{2}}\left[  b\tanh\left(  \tfrac{2abc^{2}}{1+c^{2}}\right)  +a\right]
\tfrac{c\Delta\sigma_{1}+\Delta\sigma_{2}}{\Delta d}>\tfrac{ac}{1+c^{2}%
}\left(  \tfrac{2b^{2}c^{2}}{1+c^{2}+2abc^{2}}+1\right)  \tfrac{\Delta
\sigma_{1}+\Delta\sigma_{2}}{\Delta d}>\tfrac{1}{2}\tfrac{a+b}{1+c}%
\tfrac{\Delta\sigma_{1}+\Delta\sigma_{2}}{\Delta d},
\end{equation}
and%
\begin{equation}
\tfrac{\Delta p|_{\Delta d=0}}{\Delta p|_{\Delta\sigma_{1,2}=0}}<\left(
a+b\right)  \tfrac{\Delta\sigma_{1}+\Delta\sigma_{2}}{\Delta d}%
\end{equation}
where we used%
\begin{equation}
\tfrac{x}{x+1}<\tanh x<1\text{, for }x>0.
\end{equation}

In case $\Delta\sigma_{1}\approx\Delta\sigma_{2}\approx\Delta d$, we get%
\begin{equation}
\tfrac{a+b}{1+c}<\tfrac{\Delta p|_{\Delta d=0}}{\Delta p|_{\Delta\sigma
_{1,2}=0}}<2\left(  a+b\right)  . \label{Error Bounds}%
\end{equation}

\end{document}